\documentclass[conference,twocolumn,10pt]{IEEEtran}
\ifCLASSINFOpdf
  \usepackage[pdftex]{graphicx}
\else
\fi
\usepackage{amsmath}
  \usepackage[caption=false,font=footnotesize]{subfig}
\usepackage{url}

\usepackage[linesnumbered,ruled]{algorithm2e}

\newcommand{\comment}[1]{{}}

\usepackage{amssymb}
\usepackage{xspace}

\newcommand{\complex}{\ensuremath{\mathbb{C}}\xspace}

\renewcommand{\j}{\ensuremath{\textrm{j}}}

\newcommand{\trans}{\ensuremath{^{\textsf{T}}}\xspace}
\newcommand{\ctrans}{\ensuremath{^{\textsf{H}}}\xspace}
\newcommand{\mat}[1]{\ensuremath{\mathbf{#1}}\xspace} 
\renewcommand{\vec}[1]{\ensuremath{\mathbf{#1}}\xspace} 

\newcommand{\pinv}[1]{\left({#1\ctrans} {#1}\right)^{-1}{#1\ctrans}}

\newcommand{\cgauss}[2]{\ensuremath{\mathcal{N}_{\complex} \left( {#1} , {#2} \right) }\xspace} 
\newcommand{\ev}[1]{\ensuremath{\mathbb{E}\left[{#1}\right]}\xspace}

\newcommand{\pre}{\ensuremath{\mat{F}}\xspace} 				   
\newcommand{\prebb}{\ensuremath{\mat{F}_{\textrm{BB}}}\xspace} 
\newcommand{\prerf}{\ensuremath{\mat{F}_{\textrm{RF}}}\xspace} 
\newcommand{\com}{\ensuremath{\mat{W}}\xspace} 				   
\newcommand{\combb}{\ensuremath{\mat{W}_{\textrm{BB}}}\xspace} 
\newcommand{\comrf}{\ensuremath{\mat{W}_{\textrm{RF}}}\xspace} 

\newcommand{\xbb}{\ensuremath{\mat{X}_{\textrm{BB}}}\xspace} 
\newcommand{\xrf}{\ensuremath{\mat{X}_{\textrm{RF}}}\xspace} 

\newcommand{\precbmat}{\ensuremath{\mat{A}_{\textrm{RF}}}\xspace}

\newcommand{\channel}{\ensuremath{\mat{H}}\xspace} 

\newcommand{\Nt}{\ensuremath{N_\textrm{t}}\xspace} 
\newcommand{\Nr}{\ensuremath{N_\textrm{r}}\xspace} 
\newcommand{\Na}{\ensuremath{N_{\textrm{a}}}\xspace} 

\newcommand{\Nrf}{\ensuremath{N_\textrm{RF}}\xspace} 

\newcommand{\Ns}{\ensuremath{N_\textrm{s}}\xspace} 

\newcommand{\symbvec}{\ensuremath{\vec{s}}\xspace}
\newcommand{\noisevec}{\ensuremath{\vec{n}}\xspace}
\newcommand{\rxvec}{\ensuremath{\hat{\vec{s}}}\xspace}

\newcommand{\numrays}{\ensuremath{N_{\textsf{rays}}}\xspace}
\newcommand{\numclust}{\ensuremath{N_{\textsf{clust}}}\xspace}

\newcommand{\frobenius}[1]{\ensuremath{||#1||_{\textsf{F}}}}
\newcommand{\frobeniustwo}[1]{\ensuremath{||#1||^{2}_{\textsf{F}}}}

\newcommand{\azimuth}{\ensuremath{\theta}\xspace}

\newcommand{\aoa}{\ensuremath{\theta}\xspace}
\newcommand{\aod}{\ensuremath{\phi}\xspace}

\newcommand{\rxresponsevector}{\ensuremath{\vec{a}_{\textrm{r}}(\azimuth_{i,j},\elevation_{i,j})}\xspace}
\newcommand{\txresponsevector}{\ensuremath{\vec{a}_{\textrm{t}}(\azimuth_{i,j},\elevation_{i,j})}\xspace}

\renewcommand{\rxresponsevector}{\ensuremath{\vec{a}_{\textrm{r}}}\xspace}
\renewcommand{\txresponsevector}{\ensuremath{\vec{a}_{\textrm{t}}}\xspace}

\newcommand{\snr}{\ensuremath{\textsf{SNR}}\xspace}

\newcommand{\node}[1]{\unskip\ensuremath{^{^{\left(#1\right)}}}\xspace}

\newcommand{\ctransnode}[1]{\ensuremath{^{\textsf{H}^{\left(#1\right)}}}\xspace}

\newcommand{\nullspace}[1]{\ensuremath{\textrm{null}\left(#1\right)}}

\usepackage{xspace}
\usepackage[acronym,nogroupskip,nonumberlist,nopostdot]{glossaries}
\makeglossaries

\usepackage{xcolor}

\newacronym{snr}{SNR}{signal-to-noise ratio}
\newacronym{sinr}{SINR}{signal-to-interference-plus-noise ratio}
\newacronym{inr}{INR}{interference-to-noise ratio}
\newacronym{sir}{SIR}{signal-to-interference ratio}
\newacronym{ian}{IAN}{interference as noise}
\newacronym{ber}{BER}{bit error rate}
\newacronym{pn}{PN}{pseudorandom noise}
\newacronym{bfsk}{BFSK}{binary frequency shift keying}
\newacronym{fh}{FH}{frequency-hopped}
\newacronym{fh-bfsk}{FH-BFSK}{frequency-hopped binary frequency shift keying}
\newacronym{crc}{CRC}{cyclic redundancy check}
\newacronym{isi}{ISI}{intersymbol interference}
\newacronym{dsss}{DSSS}{direct-sequence spread spectrum}
\newacronym{ofdm}{OFDM}{orthogonal frequency-division multiplexing}
\newacronym{ofdma}{OFDMA}{orthogonal frequency-division multiple access}
\newacronym{sdr}{SDR}{software-defined radio}
\newacronym{tx}{TX}{transmitter}
\newacronym{rx}{RX}{receiver}
\newacronym{fdd}{FDD}{frequency-division duplexing}
\newacronym{tdd}{TDD}{time-division duplexing}
\newacronym{fdma}{FDMA}{frequency-division multiple access}
\newacronym{tdma}{TDMA}{time-division multiple access}
\newacronym{sdma}{SDMA}{space-division multiple access}
\newacronym[plural=MPCs]{mpc}{MPC}{multipath component}
\newacronym{mui}{MUI}{multi-user interference}

\newacronym{ls}{LS}{least-squares}
\newacronym{lms}{LMS}{least mean squares}
\newacronym{rls}{RLS}{recursive least-squares}
\newacronym{mmse}{MMSE}{minimum mean square error}
\newacronym{lmmse}{LMMSE}{linear \gls{mmse}}
\newacronym{mse}{MSE}{mean square error}
\newacronym{fft}{FFT}{fast Fourier transform}
\newacronym{dft}{DFT}{discrete Fourier transform}
\newacronym{dtft}{DTFT}{discrete-time Fourier transform}
\newacronym{ctft}{CTFT}{continuous-time Fourier transform}
\newacronym{ml}{ML}{machine learning}
\newacronym[plural=NNs]{nn}{NN}{neural network}
\newacronym[plural=RNNs]{rnn}{RNN}{recurrent neural network}
\newacronym[plural=ADCs]{adc}{ADC}{analog-to-digital converter}
\newacronym[plural=DACs]{dac}{DAC}{digital-to-analog converter}
\newacronym[plural=FPGAs]{fpga}{FPGA}{field-programmable gate array}
\newacronym{evm}{EVM}{error vector magnitude}
\newacronym{psd}{PSD}{power spectral density}
\newacronym{enob}{ENOB}{effective number of bits}
\newacronym{zf}{ZF}{zero-forcing}
\newacronym{rv}{r.v.}{random variable}
\newacronym{omp}{OMP}{orthogonal matching pursuit}
\newacronym{svd}{SVD}{singular value decomposition}

\newacronym{agc}{AGC}{automatic gain control}
\newacronym{rf}{RF}{radio frequency}
\newacronym{los}{LOS}{line-of-sight}
\newacronym{nlos}{NLOS}{non-line-of-sight}
\newacronym{ple}{PLE}{path loss exponent}
\newacronym[plural=dB]{db}{dB}{decibel}
\newacronym{pa}{PA}{power amplifier}
\newacronym{lna}{LNA}{low noise amplifier}
\newacronym{cw}{CW}{continuous wave}
\newacronym{papr}{PAPR}{peak-to-average power ratio}
\newacronym{usrp}{USRP}{Universal Software Radio Peripheral}
\newacronym{irr}{IRR}{image rejection ratio}
\newacronym{lo}{LO}{local oscillator}
\newacronym{vm}{VM}{vector modulator}
\newacronym{mmwave}{mmWave}{millimeter-wave}

\newacronym{csma}{CSMA}{carrier-sense multiple access}
\newacronym{csmaca}{CSMA/CA}{carrier-sense multiple access with collision avoidance}
\newacronym{csmacd}{CSMA/CD}{carrier-sense multiple access with collision detection}
\newacronym{mac}{MAC}{medium access control}
\newacronym{phy}{PHY}{physical layer}
\newacronym{4g}{4G}{fourth generation}
\newacronym{lte}{LTE}{Long-Term Evolution}
\newacronym{4glte}{4G LTE}{\gls{4g} \gls{lte}}
\newacronym{5g}{5G}{fifth generation}
\newacronym{nr}{NR}{New Radio}
\newacronym{5gnr}{5G NR}{\gls{5g} \gls{nr}}
\newacronym{ieee}{IEEE}{Institute of Electrical and Electronics Engineers}
\newacronym{wifi}{Wi-Fi}{IEEE 802.11}
\newacronym{lan}{LAN}{local area network}
\newacronym{wlan}{WLAN}{wireless local area network}
\newacronym[plural=BSs]{bs}{BS}{base station}
\newacronym[plural=SBSs]{sbs}{SBS}{small-cell base station}
\newacronym[plural=FD-SBSs]{fdsbs}{FD-SBS}{\gls{fd}-enabled \gls{sbs}}
\newacronym[plural=MBSs]{mbs}{MBS}{macrocell base station}
\newacronym[plural=UEs]{ue}{UE}{user equipment}
\newacronym{ul}{UL}{uplink}
\newacronym{dl}{DL}{downlink}
\newacronym{qos}{QoS}{Quality of Service}
\newacronym{fcc}{FCC}{Federal Communications Commission}
\newacronym{iab}{IAB}{integrated access and backhaul}
\newacronym{fab}{FAB}{fixed access and backhaul}
\newacronym{hetnet}{HetNet}{heterogeneous network}

\newacronym{siso}{SISO}{single-input single-output}
\newacronym{mimo}{MIMO}{multiple-input multiple-output}
\newacronym{sumimo}{SU-MIMO}{single-user \gls{mimo}}
\newacronym{mumimo}{MU-MIMO}{multi-user \gls{mimo}}
\newacronym{bf}{BF}{beamforming}
\newacronym{ca}{CA}{constant amplitude}
\newacronym{ula}{ULA}{uniform linear array}
\newacronym{aoa}{AoA}{angle of arrival}
\newacronym{aod}{AoD}{angle of departure}
\newacronym{dof}{DoF}{degrees of freedom}
\newacronym{csi}{CSI}{channel state information}
\newacronym{csit}{CSIT}{\gls{csi} at the transmitter}
\newacronym{csir}{CSIR}{\gls{csi} at the receiver}

\newacronym{fd}{FD}{in-band full-duplex}
\newacronym{hd}{HD}{half-duplex}
\newacronym{si}{SI}{self-interference}
\newacronym{sic}{SIC}{self-interference cancellation}
\newacronym{soi}{SoI}{signal of interest}
\newacronym{asic}{A-SIC}{analog \gls{sic}}
\newacronym{dsic}{D-SIC}{digital \gls{sic}}
\newacronym{star}{STAR}{simultaneous transmit and receive}
\newacronym{warp}{WARP}{Wireless Open-Access Research Platform}
\newacronym{bfc}{BFC}{beamforming cancellation}
\newacronym{ipi}{IPI}{inter-panel-interference}
\newacronym{ipic}{IPIC}{inter-panel-interference cancellation}

\newacronym{qcqp}{QCQP}{quadratically-constrained quadratic programming}

\newacronym{elf}{ELF}{extremely low frequency}
\newacronym{slf}{SLF}{super low frequency}
\newacronym{ulf}{ULF}{ultra low frequency}
\newacronym{vlf}{VLF}{very low frequency}
\newacronym{lf}{LF}{low frequency}
\newacronym{mf}{MF}{medium frequency}
\newacronym{hf}{HF}{high frequency}
\newacronym{vhf}{VHF}{very high frequency}
\newacronym{uhf}{UHF}{ultra high frequency}
\newacronym{shf}{SHF}{super high frequency}
\newacronym{ehf}{EHF}{extremely high frequency}
\newacronym{thf}{THF}{tremendously high frequency}

\newacronym{wncg}{WNCG}{Wireless Networking and Communications Group}
\newacronym{linc}{LINC}{Laboratory of Informatics, Networks, and Communications}
\newacronym{ut}{UT Austin}{The University of Texas at Austin}
\newacronym{uiuc}{UIUC}{University of Illinois at Urbana-Champaign}
\newacronym{usc}{USC}{University of Southern California}
\newacronym{mit}{MIT}{Massachusetts Institute of Technology}
\newacronym{berkeley}{UC Berkeley}{University of California, Berkeley}
\newacronym{osu}{OSU}{Ohio State University}

\newcommand{\mmwave}{\gls{mmwave}\xspace}
\newcommand{\mimo}{\gls{mimo}\xspace}

\newcommand{\rf}{\gls{rf}\xspace}

\newcommand{\fg}{\gls{5g}\xspace}
\newcommand{\fd}{\gls{fd}\xspace}
\newcommand{\hd}{\gls{hd}\xspace}

\newcommand{\si}{\gls{si}\xspace}

\newcommand{\sic}{\gls{sic}\xspace}

\newcommand{\bfc}{\gls{bfc}\xspace}

\newcommand{\figref}[1]{\figurename~\ref{#1}}

\begin{document}

%
\title{Beamforming Cancellation Design for Millimeter-Wave Full-Duplex}
%
%
%


\author{%
	\IEEEauthorblockN{Ian~P.~Roberts}%
	\IEEEauthorblockA{University of Texas at Austin}%
	\and%
	\IEEEauthorblockN{Sriram~Vishwanath}%
	\IEEEauthorblockA{GenXComm, Inc., Austin, Texas, USA}%
}

\maketitle

\begin{abstract}
In recent years, there has been extensive research on \mmwave communication and on \fd communication, but work on the combination of the two is relatively lacking. \fd \mmwave systems could offer increased spectral efficiency and decreased latency while also suggesting the redesign of existing \mmwave applications. While \fd technology has been well-explored for sub-6 GHz systems, the developed methods do not translate well to \mmwave. This turns us to a method called \bfc, where the highly directional \mmwave beams are steered to mitigate \si and enable simultaneous transmission and reception in-band. In this paper, we present \bfc designs for two fully-connected hybrid beamforming scenarios, both of which sufficiently suppress the \si such that the sum spectral efficiency approaches that of a \si-free \fd system. A simulation and its results are then used to verify our designs.
\end{abstract}

%
\IEEEpeerreviewmaketitle

\glsresetall

\section{Introduction} \label{sec:introduction}

The revolution on \mmwave communication has enabled it to become a core technology of \fg cellular networks. The wide bandwidths available at \mmwave make it an attractive choice for higher rates, but challenges remain in the design of communication systems at such high carrier frequencies \cite{Andrews_Buzzi_Choi_Hanly_Lozano_Soong_Zhang_2014,Heath_Gonzalez-Prelcic_Rangan_Roh_Sayeed_2016}. Due to its high path loss, \mmwave communication necessitates the use of dense antenna arrays to provide sufficient beamforming gains to close the link and enough margin for wide-area use \cite{Heath_Gonzalez-Prelcic_Rangan_Roh_Sayeed_2016,heath_lozano}.

In \mmwave \mimo communication, it is not practical to have a dedicated \rf chain for each array element due to the associated financial cost, power consumption, size, and complexity \cite{Mendez-Rial_Rusu_Gonzalez-Prelcic_Alkhateeb_Heath_2016}. Thus, hybrid analog/digital beamforming is often employed, where the combination of a baseband (digital) beamformer and a \rf (analog) beamformer is used to achieve performance comparable to a fully-digital beamformer while reducing the number of \rf chains \cite{Heath_Gonzalez-Prelcic_Rangan_Roh_Sayeed_2016, Mendez-Rial_Rusu_Gonzalez-Prelcic_Alkhateeb_Heath_2016, Sun_Rappaport_Heath_Nix_Rangan_2014}.

First introduced about a decade ago, \fd communication enables transmission and reception using the same time-frequency resource \cite{stanford_achieving}. Such a capability offers exciting potential due to its possible doubling of spectral efficiency and the opportunity to reimagine applications that use conventional \hd techniques. To name a few, cognitive radio, relay nodes, and \gls{mac} are areas where the ability to simultaneously transmit and receive can introduce gains aside from increased spectral efficiency. When attempting to receive while transmitting, a device incurs \si which corrupts the desired receive signal if not dealt with. In order to achieve \fd, researchers have developed various \sic methods, which take advantage of the fact that a transceiver is privy to its transmit signal, enabling it to reconstruct the \si and subtract it from the corrupted signal. This leaves the desired receive signal nearly interference-free if done properly.

The majority of existing work on \fd, however, is in its application to sub-6 GHz systems. With the rise of \mmwave and its role in \fg, the extension of \fd to \mmwave is of considerable interest. However, \mmwave \fd cannot be realized by simply applying the \sic methods developed for sub-6 GHz systems. This is largely due to the numerous antennas in use, wide bandwidths, increased phase noise, and highly nonlinear components at \mmwave \cite{Satyanarayana_El-Hajjar_Kuo_Mourad_Hanzo_2019}. 

There has been introductory work on achieving \mmwave \fd using a technique known as \bfc, where the highly directional \mmwave beams are used to mitigate \si by strategic steering \cite{Satyanarayana_El-Hajjar_Kuo_Mourad_Hanzo_2019,Xiao_Xia_Xia_2017}.
When operating in a \fd fashion at \mmwave, both a transmit and receive beamformer will be in use simultaneously and will operate over the same frequencies. The goal of \bfc is to steer the transmit and receive beamformers such that they mitigate the \si while successfully transmitting to and receiving from the desired user(s). 

Without \bfc, the \si strength will likely be many orders of magnitude stronger than the incoming desired receive signal, meaning the beamformers must be designed to sufficiently mitigate the \si power while still providing sufficient gain on the desired links for successful communication. These requirements are especially true given that practical \mmwave communication is typically at low \gls{snr}.

The hybrid architectures used at \mmwave complicate \bfc due to the limitations of the analog beamformer and a desirably low number of \rf chains.
In this paper, we present two \bfc designs where \fd communication can be achieved at \mmwave under two different fully-connected hybrid beamforming scenarios.  We then present the results of simulation which validate our designs, both of which exhibit performance that can meet near ideal \fd operation. 
Additionally, we show that eigen-beamformers alone do not sufficiently suppress the \si, even with the highly directional nature of \mmwave beams.

Existing \mmwave \fd designs are limited in literature. Some designs, for example, do not consider \si at all in their work \cite{Abbas_Hamdi_2016}. The few existing \bfc designs that consider \si are iterative approaches that rely on convergence to a locally optimum solution and involve joint design across transmitter and receiver pairs \cite{Satyanarayana_El-Hajjar_Kuo_Mourad_Hanzo_2019,Jagyasi_Ubaidulla_2019}. Our design, on the other hand, is not iterative and does not rely on a joint design across nodes. Additionally, our methods consider the hybrid structure and its limitations during design whereas existing approaches assume fully-digital beamforming and decompose into hybrid approximations after the fact, not necessarily considering limitations such as phase shifter resolution.

We use the following notation. 
We use bold uppercase, \mat{A}, to represent matrices. 
We use bold lowercase, \vec{a}, to represent column vectors. 
We use $(\cdot)$\trans, $(\cdot)$\ctrans, and \frobenius{\cdot} to represent matrix transpose, conjugate transpose, and Frobenius norm, respectively. 
We use $\left[\mat{A}\right]_{i,j}$ to denote the element in the $i$th row and $j$th column of \mat{A}. 

\section{System Model} \label{sec:system-model}

\begin{figure}[!t]
	\centering
	\includegraphics[width=0.255\textwidth]{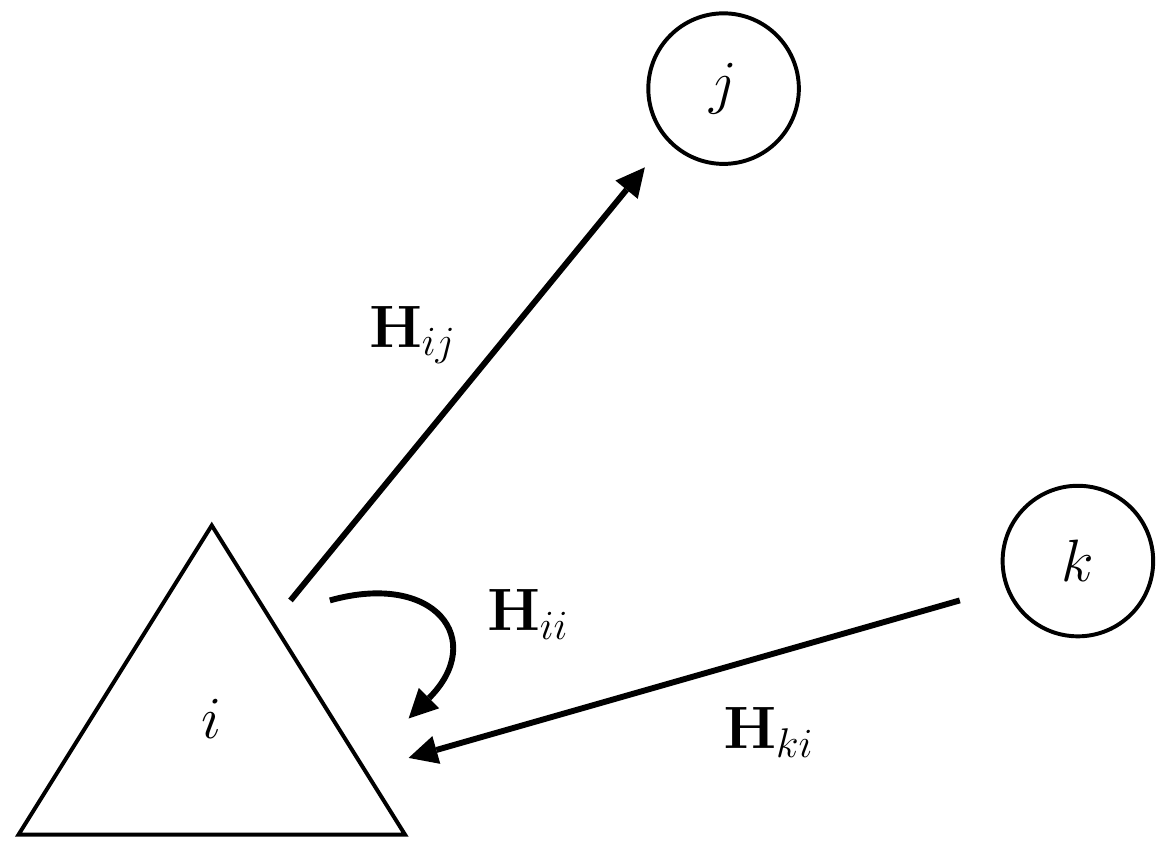}
	\caption{A \fd node $(i)$ suffering from \si as it simultaneously transmits to $(j)$ and receives from $(k)$.}
	\label{fig:network}
\end{figure}

Before presenting our design, we provide preliminary information on the system model and our assumptions.
Let us consider the \mmwave network in \figref{fig:network} where a \fd node $(i)$ is simultaneously transmitting to \hd node $(j)$ and receiving from \hd node $(k)$ in the same frequency band.

We assume $(i)$ has separate transmit and receive arrays so that it can realize two completely independent beamformers. We further assume node $(i)$ employs fully-connected hybrid beamforming  \cite{Heath_Gonzalez-Prelcic_Rangan_Roh_Sayeed_2016} where the \rf beamformers are subject to a \gls{ca} constraint whereby they have phase control but lack amplitude control.
We assume nodes $(j)$ and $(k)$ can perfectly realize a fully-digital beamformer via hybrid beamforming (e.g., using \cite{wei-yu}). Thus, we can ignore the hybrid architecture at $(j)$ and $(k)$.

We let the number of transmit antennas, $\Nt$, and the number of receive antennas, $\Nr$, be the same at all nodes for convenience of notation and in analysis. Since we are focusing on the hybrid architecture at $(i)$, we let $\Nrf$ be the number of \rf chains at its transmit array and at its receive array. At each node, we impose a per-symbol power constraint along with uniform power allocation across streams such that the Frobenius norm of each stream's precoder must be $\sqrt{\Nt}$.

Let $\prebb\node{i}$, $\prerf\node{i}$, $\comrf\node{i}$, and $\combb\node{i}$ denote the baseband precoder, \rf precoder, \rf combiner, and baseband combiner, all at $(i)$, respectively. 
Abiding by the \gls{ca} constraint, the entries of $\prerf\node{i}$ and $\comrf\node{i}$ are required to have unit magnitude. 
Let $\pre\node{k}$ be the precoder at $(k)$ and $\com\node{j}$ be the combiner at $(j)$.

Let $\channel_{ki}$ be the $\Nr \times \Nt$ channel from $(k)$ to $(i)$.
Let $\channel_{ij}$ be the $\Nr \times \Nt$ channel from $(i)$ to $(j)$.
Let $\channel_{ii}$ be the $\Nr \times \Nt$ \si channel from the transmit array of $(i)$ to its own receive array.
Let $\symbvec\node{i} \sim \cgauss{\vec{0}}{\mat{I}}$ and $\symbvec\node{j} \sim \cgauss{\vec{0}}{\mat{I}}$ be the $\Ns \times 1$ symbol vectors intended for $(i)$ and $(j)$, respectively, where we have assumed the same number of streams is being sent on both links (and that each node has support of \Ns streams).
Let $\noisevec\node{i} \sim \cgauss{\vec{0}}{\mat{I}}$ and $\noisevec\node{j} \sim \cgauss{\vec{0}}{\mat{I}}$ be $\Nr \times 1$ noise vectors at the receive arrays of $(i)$ and $(j)$, respectively.

We define the \gls{snr} for a given link as accounting for transmit power and the link's path loss. Let $\snr_{ki}$ and $\snr_{ij}$ be the \gls{snr} from $(k)$ to $(i)$ and from $(i)$ to $(j)$, respectively. Let $\snr_{ii}$ be the \gls{snr} of the \si.

Let us focus our attention on the \fd node $(i)$. We can represent its estimated receive symbol $\rxvec\node{i}$ as follows, where \eqref{eq:term-desired} is the desired term, \eqref{eq:term-si} is the resulting \si term, and \eqref{eq:term-noise} is the resulting noise term.
\begin{align}
\rxvec\node{i} &= \sqrt{\snr_{ki}} \ \combb\ctransnode{i} \comrf\ctransnode{i} \channel_{ki} \pre\node{k} \symbvec\node{i} \label{eq:term-desired} \\
&+ \sqrt{\snr_{ii}} \ \combb \ctransnode{i} \comrf\ctransnode{i} \channel_{ii} \prerf\node{i} \prebb\node{i} \symbvec\node{j} \label{eq:term-si} \\
&+ \combb\ctransnode{i} \comrf\ctransnode{i} \noisevec\node{i} \label{eq:term-noise}
\end{align}
As indicated by \eqref{eq:term-si}, the transmitted signal from $(i)$ intended for $(j)$ traverses through the \si channel $\channel_{ii}$ and is received by the combiner at $(i)$ being used for reception from $(k)$.  

The estimated receive symbol at $(j)$ from $(i)$ is
\begin{align}
\rxvec\node{j} &= \sqrt{\snr_{ij}} \ \com\ctransnode{j} \channel_{ij} \prerf\node{i} \prebb\node{i} \symbvec\node{j} + \com\ctransnode{j} \noisevec\node{j} \label{eq:term-j},
\end{align}
where $(j)$ does not encounter any \si since it is a \hd device. Additionally, we assume that there is no adjacent user interference between $(j)$ and $(k)$. We justify this assumption with the high path loss and directionality of transmission and reception at \mmwave.

For the desired link channels, $\channel_{ij}$ and $\channel_{ki}$, we assume sufficient far-field conditions have been met. 
We employ the extended Saleh-Valenzuela \mmwave channel representation as shown in \eqref{eq:clustered-channel-model}. In this model, an $\Nr\times\Nt$ channel is the sum of the contributions from \numclust scattering clusters, each of which contributes \numrays propagation paths \cite{Mendez-Rial_Rusu_Gonzalez-Prelcic_Alkhateeb_Heath_2016}. 
\begin{equation} \label{eq:clustered-channel-model}
\mat{H} = \sqrt{\frac{\Nt \Nr}{\numrays \numclust}} \sum_{m=1}^{\numclust} \sum_{n=1}^{\numrays} \beta_{m,n} \rxresponsevector(\aoa_{m,n}) \txresponsevector\ctrans(\aod_{m,n})
\end{equation}
In \eqref{eq:clustered-channel-model}, $\rxresponsevector(\aoa_{m,n})$ and $\txresponsevector(\aod_{m,n})$ are the antenna array responses 
at the receiver and transmitter, respectively, for ray $n$ within cluster $m$ which has some \gls{aoa}, $\aoa_{m,n}$, and \gls{aod}, $\aod_{m,n}$. Each ray has a random gain $\beta_{m,n} \sim \cgauss{0}{1}$. The coefficient outside the summations is used to ensure that $\ev{ \frobeniustwo{ \mat{H} }} = \Nt\Nr$. Note that we have considered a single angular dimension (e.g., azimuth) for the \glspl{aoa} and \glspl{aod} for simplicity and to agree with our use of \glspl{ula} in simulation.

When $(i)$ transmits to $(j)$ while receiving from $(k)$, there will almost certainly be some undesired leakage from the transmit array to the receive array that corrupts the desired receive signal. The strength of this leakage is a function of the  precoder at $(i)$, the combiner at $(i)$, and the channel between them. The transmit and receive arrays at $(i)$ will be relatively close together, meaning the far-field condition is not met and near-field effects must be considered. Unfortunately, the \mmwave \si channel is much less understood since research and measurements at \mmwave have primarily focused on more practical far-field scenarios. Therefore, we have turned to one of the few attempts at characterizing the \mmwave \si channel \cite{Satyanarayana_El-Hajjar_Kuo_Mourad_Hanzo_2019}, where it is decomposed in a Rician fashion containing a near-field \gls{los} component and a far-field component stemming from \gls{nlos} reflections. Explicitly, we model the \si as 
\begin{align} \label{eq:si-channel-rice}
	\channel_{ii} &= \sqrt{\frac{\kappa}{\kappa + 1}} \channel^{\textrm{LOS}}_{ii} + \sqrt{\frac{1}{\kappa + 1}} \channel^{\textrm{NLOS}}_{ii}
\end{align}
where $\kappa$ is the Rician factor. The entries of the \gls{los} contribution are modeled as \cite{spherical-wave-mimo}
\begin{align} \label{eq:si-channel-los}
	\left[\channel^{\textrm{LOS}}_{ii}\right]_{m,n} &= \frac{\rho}{r_{m,n}}\exp \left(-\j 2 \pi \frac{r_{m,n}}{\lambda_{c}} \right)
\end{align}
where $\rho$ is a normalization constant such that $\ev{\frobeniustwo{\channel_{ii}}} = \Nt\Nr$ and $r_{m,n}$ is the distance from the $m$th element of the transmit array to the $n$th element of the receive array. We leave the definition of $r_{m,n}$ to \cite{spherical-wave-mimo} for space considerations. For the \gls{nlos} \si channel, we use the model in \eqref{eq:clustered-channel-model}.

\section{Proposed Beamforming Cancellation Designs} \label{sec:bfc}

In order for $(i)$ to simultaneously transmit and receive in-band, we seek to mitigate the \si by design of its transmit and/or receive beamformers. Performance on the desired links must be maintained to some degree or else transmission and reception will be poor. 
We assume $(i)$ and $(j)$ have perfect \gls{csi} of $\channel_{ij}$ and that $(k)$ and $(i)$ have perfect \gls{csi} of $\channel_{ki}$. We assume $(i)$ also has perfect \gls{csi} of its \si channel, $\channel_{ii}$. 

To begin the design, we set $\pre\node{k}$ and $\com\node{j}$ to be the so-called eigen-beamformers for their respective channels as follows. Taking the \gls{svd} of $\channel_{ki}$ and of $\channel_{ij}$, we get
\begin{align}
	\channel_{ki} &= \mat{U}_{ki} \mat{\Sigma}_{ki} \mat{V}\ctrans_{ki}
 \label{eq:svd-hki} \\
 \channel_{ij} &= \mat{U}_{ij} \mat{\Sigma}_{ij} \mat{V}\ctrans_{ij} \label{eq:svd-hij}
\end{align}
where $ \mat{\Sigma}_{ki}$ and $ \mat{\Sigma}_{ij}$ contain decreasing singular values along their diagonals. Setting $\pre\node{k}$ and $\com\node{j}$ as the singular vectors corresponding to the $\Ns$ strongest eigenchannels, we have
\begin{align}
\pre\node{k} &= \left[ \mat{V}_{ki} \right]_{:,0:\Ns-1} \label{eq:pre-k} \\
\com\node{j} &= \left[ \mat{U}_{ij} \right]_{:,0:\Ns-1} \label{eq:com-j}
\end{align}
where we use $\left[\cdot\right]_{:,0:\Ns-1}$ to denote selecting the \Ns leftmost columns. Abiding by our power constraint, \eqref{eq:pre-k} is normalized such that its columns have Frobenius norm $\sqrt{\Nt}$. 
We have assumed $(k)$ and $(j)$ can perfectly realize these fully-digital beamformers.
Furthermore, having assumed perfect \gls{csi} implies that $(i)$ has knowledge of $\pre\node{k}$ and $\com\node{j}$.

Since $(i)$ employs hybrid beamforming, constructing a desired beamformer occurs in an analog stage and a digital stage. The analog beamformer and the number of \rf chains impose limitations on the accuracy of the hybrid construction. For this reason, we explore two scenarios based on the hybrid architecture in use. The first is a less practical case when the hybrid beamformers have infinite precision phase shifters and twice the number of \rf chains as streams, i.e., $\Nrf = 2\Ns$. In the second scenario, we consider a more practical hybrid architecture where $2\Ns > \Nrf \geq \Ns$ and the phase shifters have finite resolution. In both cases, we assume the analog beamformers abide by the \gls{ca} constraint and that fully-connected hybrid beamforming is employed.

\subsection*{Case A: $\Nrf = 2\Ns$ \& Infinite Resolution Phase Shifters} \label{sec:scenario-1}
In this scenario, we assume that the phase shifters have infinite precision and that there are twice as many \rf chains as there are streams.
Given these assumptions, we reference the work of \cite{wei-yu} where it was shown that having $\Nrf = 2\Ns$ and infinite precision phase shifters enables perfect hybrid decomposition of any fully-digital beamformer. This alleviates our \bfc design in Case A of any hybrid architectural constraints and allows us to decompose a fully-digital design without penalty.

To begin our design, we first choose the fully-digital combiner at $(i)$, $\com\node{i}$, to be the eigen-combiner for its link with $(k)$. Using \eqref{eq:svd-hki}, we get 
\begin{align}
\com\node{i}  &= \left[ \mat{U}_{ki} \right]_{:,0:\Ns-1}. \label{eq:hki-left-singular}
\end{align}
Choosing to fix $\com\node{i}$ allows us to consider the \textit{effective} \si channel, which is the portion of the \si channel that gets received by the combiner at $(i)$ as it receives from $(k)$. This effective \si channel is then
\begin{align}
\com\ctransnode{i} \channel_{ii}. \label{eq:effective-si-ideal}
\end{align}

Ideally, our choice for $\pre\node{i}$ would be such that it contributes no \si while transmitting to $(j)$. Put simply, we desire 
\begin{align}
	\com\ctransnode{i} \channel_{ii} \pre\node{i} = \mat{0}. \label{eq:no-si-goal}
\end{align}
Setting $\pre\node{i}$ to the eigen-precoder, however, would very likely violate \eqref{eq:no-si-goal}, introducing \si, corrupting the signal from $(k)$, and degrading the spectral efficiency of its link with $(k)$. The eigen-precoder and eigen-combiner would satisfy \eqref{eq:no-si-goal} only if they happened to be orthogonal.

Fixing $\com\node{i}$, for \eqref{eq:no-si-goal} to hold, \pre\node{i} must lie in the null space of $\com\ctransnode{i} \channel_{ii}$. An alternate interpretation as to why we have chosen to consider the effective \si channel rather than the entire \si channel is as follows. The \si channel model we have employed in \eqref{eq:si-channel-rice} and \eqref{eq:si-channel-los} is full rank, leaving no dimensions for the null space by the rank-nullity theorem. This has inspired us to more appropriately consider the effective \si channel in \eqref{eq:effective-si-ideal} rather than solely $\channel_{ii}$, leaving a null space with many dimensions for $\pre\node{i}$ to live. In short, we only need to be concerned with the portion of \si that is \textit{received}.

Since it is likely that the eigen-precoder will not naturally lie in the null space of the effective \si channel, we seek to project it as follows.
Let 
$\mat{B} = \nullspace{\com\ctransnode{i} \channel_{ii}}$ 
be a matrix whose columns are a basis of the null space of the effective \si channel. The matrix $\mat{B}$ can be found in multiple ways, a simple one being comprised of the right singular vectors of the effective \si channel that correspond to singular values of zero. The projection matrix for the column space of $\mat{B}$ is then
\begin{align}
\mat{P} = \mat{B} \pinv{\mat{B}}
\end{align}
which is used to project the eigen-precoder onto $\mat{B}$ by
\begin{align}
\pre\node{i} &= \mat{P} \left[ \mat{V}_{ij} \right]_{:,0:\Ns-1} \label{eq:pre-final-1}
\end{align}
where $\mat{V}_{ij}$ are the right singular vectors from \eqref{eq:svd-hij}. Abiding by our power constraint, \eqref{eq:pre-final-1} is normalized such that its columns have Frobenius norm $\sqrt{\Nt}$. 

Given this scenario's hybrid assumptions, we invoke the method in \cite{wei-yu} for determining the analog and digital beamformers that perfectly construct the desired fully-digital beamformers $\pre\node{i}$ and $\com\node{i}$ in a hybrid fashion.

\subsection*{Case B: $2\Ns > \Nrf \geq \Ns$ \& Finite Resolution Phase Shifters} \label{sec:scenario-2}
In this scenario, we consider more practical implementation conditions, where the number of \rf chains is less than twice the number of streams, $2\Ns > \Nrf \geq \Ns$, and the phase shifters have finite resolution.
Our design does not rely on a specific phase shifter resolution, though higher resolution phase shifters will naturally yield better performance.

A popular and effective hybrid decomposition algorithm for such a scenario uses \gls{omp} to decompose a desired fully-digital beamformer into its hybrid counterpart by capturing the analog limitations in an analog beamforming codebook \cite{omp-heath}. We cannot directly apply this \gls{omp}-based decomposition to the fully-digital solution found in Case~A, however, since the hybrid approximation of \pre\node{i} will not necessarily lie in the null space of the effective \si channel---orthogonality is likely lost during hybrid approximation. Instead, our method for this case projects \pre\node{i} into the null space of the effective \si channel \textit{during} hybrid decomposition, rather than before as in Case~A. Our design is as follows.

We first choose $\com\node{i}$ to be the eigen-combiner for its link with $(k)$ as shown in \eqref{eq:hki-left-singular}. Since we are considering the aforementioned practical limitations, we must approximate this fully-digital beamformer by its hybrid counterpart. We choose to perform this hybrid decomposition before proceeding rather than later as will be apparent.
The hybrid approximation of $\com\node{i}$ can be done using the \gls{omp}-based approach, for example, though our design does not require a particular hybrid decomposition method for the combiner. Following hybrid approximation, the effective combiner at $(i)$ is $\comrf\node{i} \combb\node{i}$.

We begin the design of \pre\node{i} by initializing it to the eigen-precoder
\begin{align}
\pre\node{i} &= \left[ \mat{V}_{ij} \right]_{:,0:\Ns-1}. \label{eq:init-pre-eigen}
\end{align}
Let $\precbmat$ be a matrix whose columns are valid analog beamforming vectors. 
We invoke a conventional \gls{omp}-based hybrid approximation of $\pre\node{i}$ using $\precbmat$ which returns an analog precoder \xrf and a digital precoder \xbb whose product approximates the eigen-precoder and meets the hybrid constraints. 

To ensure that there is no leakage of a transmission into the receive array and since the analog precoder is limited by its constraints, we project the digital precoder into the null space of the effective \si channel \eqref{eq:effective-si-practical}. Thus, we set the analog beamformer as $\prerf\node{i} = \xrf$ and focus only on the digital beamformer.

Having set the hybrid combiner to $\comrf\node{i}\combb\node{i}$ and the analog precoder to $\prerf\node{i}$, the effective \si channel becomes
\begin{align}
\combb\ctransnode{i} \comrf\ctransnode{i} \channel_{ii} \prerf\node{i}. \label{eq:effective-si-practical}
\end{align}
Let $\mat{C} = \nullspace{\combb\ctransnode{i} \comrf\ctransnode{i} \channel_{ii} \prerf\node{i}}$ be a matrix whose columns are a basis of the null space of the effective \si channel in \eqref{eq:effective-si-practical}. The projection matrix for $\mat{C}$ is
\begin{align}
\mat{Q} = \mat{C} \pinv{\mat{C}}.
\end{align}
which is used to project the digital precoder \xbb onto the column space of $\mat{C}$ by
\begin{align}
	\prebb\node{i} &= \mat{Q} \xbb. \label{eq:projecting-digital}
\end{align}
The effective precoder at $(i)$ is then
\begin{align}
\pre\node{i} &= \prerf\node{i} \prebb\node{i} = \xrf \mat{Q} \xbb \label{eq:composite-precoder-practical}
\end{align}
which is normalized such that its columns have Frobenius norm $\sqrt{\Nt}$.
This concludes our design, which we validate in the next section using simulation. In both designs, for Case A and Case B, our transmit precoder is designed such that reception from $(k)$ is unaffected by transmission to $(j)$.

\section{Simulation and Results} \label{sec:simulation-results}

In order to verify our proposed method, a simulation was developed in MATLAB. 
We denote the carrier wavelength as $\lambda_{\textrm{c}}$. We assume all antenna arrays to be \glspl{ula} with element spacing $\lambda_{\textrm{c}} / 2$. Each array element is assumed to be isotropic. The array response vector can be formulated as $\vec{a}(\azimuth) = \left[1 \ e^{\j \varphi(\azimuth)} \ e^{\j 2 \varphi(\azimuth)} \ \dots \ e^{\j (\Na-1) \varphi(\azimuth)} \right]\trans \label{eq:ula-response-vector}$
where $\Na$ is the number of elements in the array and 
$\varphi(\azimuth) = \frac{2 \pi}{\lambda_{\textrm{c}}} \frac{\lambda_{\textrm{c}}}{2} \cos(\azimuth) \label{eq:phi-def}$
is the phase shift between elements.

For generating the channels $\channel_{ki}$ and $\channel_{ij}$, we use a random number of clusters and rays per cluster uniformly distributed on $[1,6]$ and $[1,10]$, respectively. Within a given cluster, each ray's \gls{aod} and \gls{aoa} are Laplacian distributed sharing a mean which is uniformly distributed on $[0,\pi]$ and having standard deviation of $0.2$.

The \si Rician factor is chosen to be $\kappa = 30$ dB, since we assume the \gls{los} \si would dominate in practice. We assume the transmit and receive arrays at $(i)$ are separated by a distance $10 \lambda_{\textrm{c}}$ and an angle $\pi/6$. For generating $\channel^{\textrm{NLOS}}_{ii}$, we do so in the same fashion as the desired links $\channel_{ki}$ and $\channel_{ij}$ but having a random number of clusters and rays uniformly distributed on $[1,3]$ and $[1,3]$, since we assume there would be less scattering due to the close proximity. The \si \gls{snr} is set to be $\snr_{ii} = 120$ dB where we have imagined a practical scenario having a \si power of $30$ dBm at the receive array and a noise floor of $-90$ dBm. We let the link \glspl{snr} from $(k)$ to $(i)$ and from $(i)$ to $(j)$ to be equal for the sake of interpreting the results, i.e., $\snr_{ki} = \snr_{ij} = \snr$.




\subsection*{Results of Case A}
\begin{figure}[!t]
	\centering
	\includegraphics[width=0.46\textwidth]{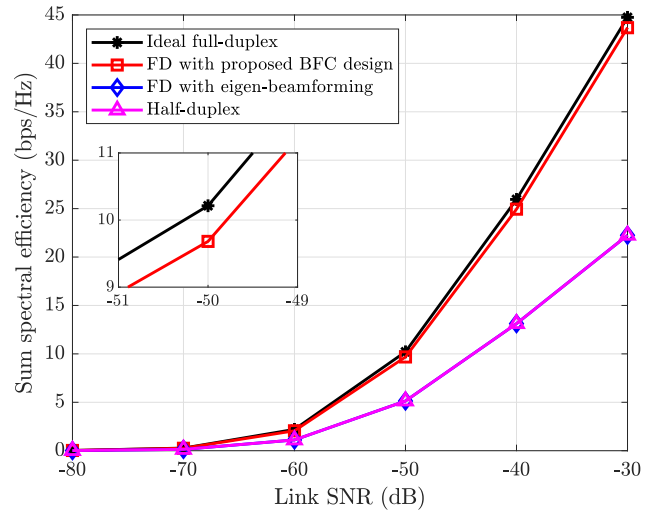}
	\caption{Sum spectral efficiencies achieved with infinite precision phase shifters when $\Ns = 3$, $\Nrf = 6$, and $\Nt = \Nr = 16$.}
	\label{fig:results-1-16}
\end{figure}

\begin{figure}[!ht]
	\centering
	\includegraphics[width=0.46\textwidth]{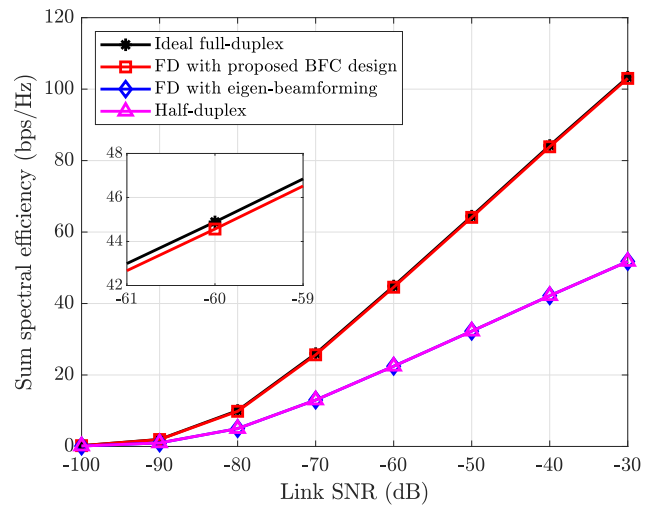}
	\caption{Sum spectral efficiencies achieved with infinite precision phase shifters when $\Ns = 3$, $\Nrf = 6$, and $\Nt = \Nr = 64$.}
	\label{fig:results-1-64}
\end{figure}

Having infinite precision phase shifters and $\Nrf = 2\Ns$, we can perfectly replicate our design for Case A using hybrid beamforming. In \figref{fig:results-1-16} and \figref{fig:results-1-64}, we compare various spectral efficiencies for $\Nt = \Nr = 16$ and $\Nt = \Nr = 64$, respectively, where $\Ns = 3$ in both. 

When using its eigen-precoder and eigen-combiner, $(i)$ will almost certainly collect some portion of \si. Since the \si is extremely strong relative to a desired receive signal, even a small fraction of \si overwhelms the desired receive signal, making successful reception from $(k)$ nearly impossible. Transmission to $(j)$ is not impacted since we are using the eigen-precoder and $(j)$ is unaffected by the \si at $(i)$.
This results in a spectral efficiency to $(j)$ equivalent to its \hd spectral efficiency and a spectral efficiency from $(k)$ being practically zero due to the dominating \si. Thus, the sum spectral efficiency using eigen-beamforming at $(i)$ is nearly identical to the \hd case where only a single node, $(i)$ or $(k)$, transmits. These results indicate that eigen-beamforming is not sufficient for \fd operation even at \mmwave where the beams are highly directional.

With our \bfc design, however, the sum spectral efficiency hugs the ideal \fd spectral efficiency curve quite closely. Ideal \fd operation is simply the case when the degrading effects of \si are not present, allowing both \hd links to operate at their maximum rates simultaneously.
By design, we completely eliminate the received \si, allowing the link from $(k)$ to achieve a spectral efficiency equal to its \hd spectral efficiency even while $(j)$ is simultaneously being served. 
The spectral efficiency of the link from $(i)$ to $(j)$ is slightly degraded compared to during \hd operation due to $\pre\node{i}$ deviating from the eigen-precoder as it is projected into the null space of the effective \si---this is the only cost of operating in \fd with our \bfc design.
The ability to perfectly recreate our fully-digital beamformers in a hybrid fashion in this scenario allows us to attribute any deviation from ideal \fd performance to our design and not to the hybrid approximation.

When $64$ antennas are in use, there are two important factors at play. The first is that higher spectral efficiencies can be achieved at a given link \gls{snr} compared to with $16$ antennas, of course, due to the increased beamforming gain. The second factor is that, with more antennas but constant $\Ns$, our null space dimension grows, allowing us to more accurately approximate our eigen-precoder while constraining it to live in the null space of the effective \si.

\subsection*{Results of Case B}
For the scenario where we have finite precision phase shifters and $2\Ns > \Nrf \geq \Ns$, we have used \gls{omp}-based hybrid approximation which uses a codebook for choosing analog beamformers that satisfy the \gls{ca} constraint and phase shifter resolution.
For simulation, we have used a \gls{dft} analog beamforming codebook, which does not contain anywhere near every possible analog beamformer meaning it could result in sub-optimality. However, since we have taken this scenario to be a practical one, using the \gls{dft} codebook is deemed a suitable choice for its computational convenience and sufficient performance with \glspl{ula}. It is worth noting that this choice implicitly assumes that our phase shifters have a resolution of at least $2\pi/\Na$, where $\Na$ is the number of antennas at the beamformer. In general, an appropriate beamforming codebook should be chosen based on the particular analog beamformer's limitations.
Given this sub-optimality, we will observe some loss during projection into the null space in addition to the loss during hybrid approximation. We have observed, however, that the loss during hybrid approximation is fairly insignificant, especially with increasing \Nrf. 

\begin{figure}[!t]
	\centering
	\includegraphics[width=0.46\textwidth]{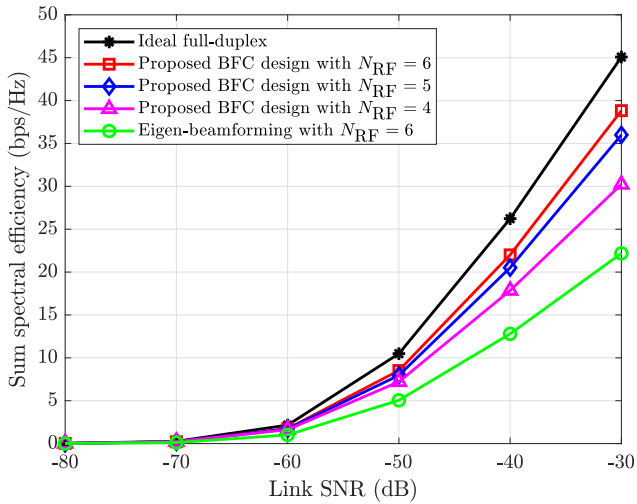}
	\caption{Sum spectral efficiency for varying number of \rf chains with finite resolution phase shifters when $\Ns = 3$ and $\Nt = \Nr = 16$.}
	\label{fig:results-2-16}
\end{figure}

\begin{figure}[!t]
	\centering
	\includegraphics[width=0.46\textwidth]{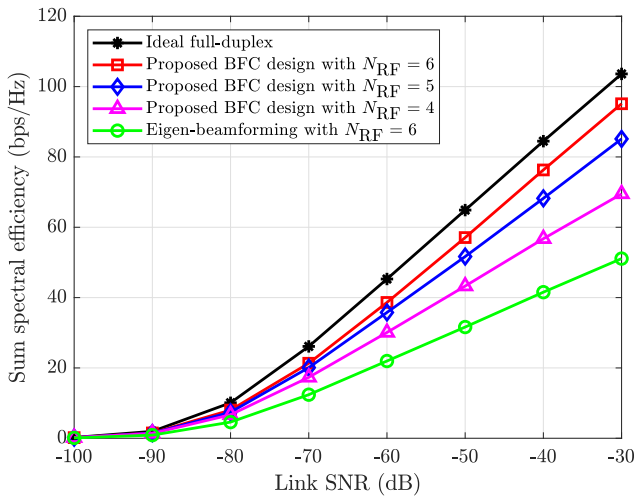}
	\caption{Sum spectral efficiency for varying number of \rf chains with finite resolution phase shifters when $\Ns = 3$ and $\Nt = \Nr = 64$.}
	\label{fig:results-2-64}
\end{figure}

As seen in \figref{fig:results-2-16} and \figref{fig:results-2-64}, our \bfc design incorporating the hybrid architectural constraints performs quite well, especially for an increasing number of \rf chains. In both figures, we let $\Ns = 3$ and vary the number of \rf chains. 
With $\Nrf \geq 4$, we see that our \bfc design outperforms the hybrid-approximated eigen-precoder having $\Nrf = 6$, netting a higher sum spectral efficiency. This increase in sum spectral efficiency is attributed to increased performance on the link from $(i)$ to $(j)$, primarily due to more closely approaching the eigen-precoder while projecting it into the null space of the effective \si channel. 
In other words, increasing the number of \rf chains allows us to suppress the \si while better transmitting to $(j)$ by offering more dimensions to the digital precoder $\prebb\node{i}$. 
Our observations attribute only marginal gain to improved hybrid approximation as \Nrf increases---the spectral efficiency gains are primarily due to improved \bfc.

\section{Conclusion} \label{sec:conclusion}


In this paper, we have presented beamforming designs for \mmwave \fd, enabling simultaneous transmission and reception in-band. 
Accounting for the hybrid architectures used at \mmwave, we have proposed two designs: one for a more design-favorable scenario and another for a more practical one. In both, we abide by a \gls{ca} constraint at the analog beamformers and employ a fully-connected hybrid architecture. 
We show that a device using our designs can achieve significant spectral efficiency gains approaching that of ideal \fd operation.
Our results indicate that \mmwave \fd applications can be enabled by our designs, offering increased spectral efficiency with the ability to transmit while receiving, even under the constraints imposed by hybrid beamforming.

\section*{Acknowledgments}
This work was supported by the National Science Foundation under grants 1731754 and 1559997.

\bibliography{refs_2}

\end{document}